\documentclass[aps, prb, twocolumn,10pt, floatfix]{revtex4-2}
\usepackage{amsmath}
\usepackage{amssymb}
\usepackage{graphicx}
\usepackage{textcomp}
\usepackage[normalem]{ulem}
\usepackage{siunitx}
\usepackage[T1]{fontenc}
\usepackage[percent]{overpic}

\newcommand{\mi}{\textmu}
\newlength{\kol}
\begin{document}
\title{Low-Divergence Quasi-Gaussian Emission at Watt-Level Power from a~Large-Diameter Ring-Aperture VCSEL}

\author{Marta~Więckowska$^1$, Luke~Graham$^2$, James~Guenter$^2$, 
Jim~Tatum$^2$, Freddie~Castillo$^2$, Karolina~Olucha$^3$, 
Justyna Maleszyk$^3$,
Magdalena~Marciniak$^1$, Michał Dobrski$^1$, Tomasz~Czyszanowski$^1$, Michał~Wasiak$^1$}
\affiliation{$^1$ Institute of Physics, Lodz University of Technology, Wólczańska 217/221, 91-005 Łódź, Poland}
\affiliation{$^2$ Dallas Quantum Devices, 1575 Redbud Suite 212, McKinney, TX USA 75069}
\affiliation{$^3$ Łukasiewicz Research Network -- Institute of Microelectronics and Photonics, al.~Lotników 32/46, 02-668 Warsaw, Poland}
\begin{abstract}
The far-field emission of large-area vertical-cavity surface-emitting lasers (VCSELs) is commonly associated with multimode, high-divergence beam profiles, limiting applicability in high-brightness free-space systems. We investigate angular emission characteristics of a \SI{1}{\milli\meter}-diameter ring-aperture watt-class VCSEL and establish a theoretical framework capturing the formation of its far-field radiation patterns. Modeling the near field as an azimuthally modulated ring distribution and evaluating the far field within the Fresnel approximation, we demonstrate that a quasi-Gaussian far-field profile emerges from combined lower-order azimuthal modes, even in a highly multimode cavity.

Experimentally, we observe a current-driven transition of the far-field distribution from a high-divergence ring at low injection levels to a narrow central beam at elevated currents. At high drive currents, the emission approaches a near-Gaussian profile with a full width at half maximum of \SI{8}{\degree}, while maintaining watt-class output power. Angle-resolved spectroscopy associates the central emission with longer-wavelength, lower-order modes, whereas the outer ring originates from shorter-wavelength, higher-order contributions. Combined with electroluminescence measurements and wavelength-dependent photon lifetime analysis, these results demonstrate that spectral and angular emission are determined by the interplay between wavelength-dependent material gain and angle-dependent cavity losses. This approach establishes a general framework for controlling beam divergence and modal content in large-area VCSELs, enabling high-power operation with near-Gaussian, low-divergence beam profiles.
\end{abstract}
\maketitle
\section{Introduction}
Vertical-cavity surface-emitting lasers (VCSELs) have emerged over the last thirty years as one of the most important compact light sources, valued for their small footprint, record-low threshold currents in the sub-milliampere range~\cite{wu2025sub}, high modulation speed reaching 30\,GHz~\cite{yang202130}, and high energy conversion efficiency exceeding 70\% in multiple tunnel junction devices~\cite{xiao2024multi}. These characteristics have made them indispensable in high-speed optical interconnects and short-reach data communication, as well as in a wide range of sensing applications. In recent years, applications such as facial recognition, structured-light illumination, time-of-flight depth ranging, and 3D imaging in consumer devices ranging from smartphones to robotics have driven production volumes of VCSEL chips and modules into the billions annually. Today, their use is rapidly expanding into new sectors, including automotive LiDAR, high-performance computing, virtual and augmented reality, industrial processing and heating, and even medical aesthetics~\cite{pan2024harnessing,koyama2014advances}.

In conventional VCSELs with circular oxide apertures, increasing the aperture diameter increases the emitted optical power. Initially, for small aperture diameters, the maximum output power scales proportionally with the aperture area (quadratic dependence on the aperture radius). However, with increasing aperture, the lateral current distribution in the active region becomes increasingly non-uniform, leading to local carrier accumulation at the aperture perimeter~\cite{Degen1999,Grabherr1999}. This non-uniform injection leads to enhanced self-heating and efficiency degradation, since radiative recombination is weak in the central region of the aperture due to carrier depletion. Additionally, lateral current inhomogeneity promotes the excitation of higher-order transverse modes near the aperture perimeter, leading to multimode emission and degraded beam quality~\cite{Degen1999,Grabherr1999}. As a result, the output power exhibits a linear rather than quadratic dependence on the aperture radius.

To overcome these limitations, high-power VCSEL systems are typically implemented as two-dimensional arrays of small emitters, which retain nearly linear power scaling with the total aperture area, enabling optical outputs in the hundred-watt range~\cite{seurin2008high} while maintaining effective thermal management~\cite{Liu2024}. However, densely packed arrays introduce their own challenges, including complex current distribution, thermal crosstalk, increased beam divergence—typically above $10^\circ$ as a compromise with emitted power~\cite{Zhang2012,Okur2019}—and reduced overall brightness due to the incoherent superposition of emission from multiple elements~\cite{Pan2022}.
\begin{figure}[tbh]
  \includegraphics[width=\kol]{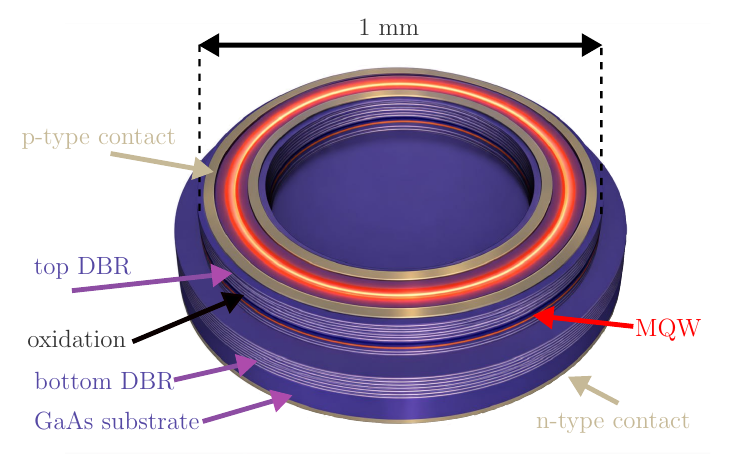}
  \caption{Schematic illustration of a coherent ring VCSEL with large-area, ring-shaped apertures.}
  \label{fig:ring_vcsel}
\end{figure}
In the past, several device architectures were proposed to balance the need for low-divergence emission with the high-power output characteristic of two-dimensional VCSEL arrays. The distributed-ring contact design demonstrated watt-class continuous-wave operation with improved emission uniformity and a beam divergence of $\sim12^\circ$. Multi-ring-shaped apertures further alleviated current crowding and provided enhanced beam profiles, albeit with output powers limited to a few hundred milliwatts~\cite{Hao2011}. The reticular electrode structure extended these efforts by improving thermal stability~\cite{Feng2015}, while ring close-packing contacts enabled powers above 1\,W with narrow divergence angles increasing from $\sim3^\circ$ at threshold to $\sim15^\circ$ at rollover, yielding near-Gaussian far-field distributions~\cite{Li2020}.

An attractive approach in this context is the use of VCSELs with large-area ring-shaped apertures (see Figure~\ref{fig:ring_vcsel} for a schematic). Such a large-area architecture contributes to the reduction of beam divergence and enables high-power continuous-wave operation from a single device. The distributed current injection around a ring aperture of relatively small width mitigates carrier distribution inhomogeneity. The aperture's circular symmetry and small radial width reduce thermal gradients across the aperture and therefore decrease the associated wavelength spread compared to conventional large-area VCSELs and VCSEL arrays. A similar ring architecture has been reported in earlier studies~\cite{Shi2005,Yan2025}, but the aperture diameter did not exceed $100\,$\mi m, which limited the output power to only a few tens of milliwatts.

Building on this foundation, we demonstrate in this work that a lithographically defined large coherent-ring aperture VCSEL (see Figure~\ref{fig:ring_vcsel}) achieves a record balance between beam quality and power, combining a Gaussian far-field with an $8^\circ$ divergence—enabled by complex mode interactions—with continuous-wave output powers exceeding 1\,W. These exceptional characteristics are further complemented by a wall-plug efficiency of 30\%, thereby surpassing the limitations of conventional large-area VCSEL architectures and establishing a pathway toward high-brightness sources for applications such as LiDAR, free-space optical communications, and optical pumping.
\section{Theoretical determination of far fields}\label{sec:theo}
In order to calculate far-fields patterns for the ring-aperture VCSELs,
we use the Fresnel approximation. It allows us to calculate
the far field of the laser based on an assumed near-field distribution
(i.e~on the surface of the laser aperture). This approximation is valid
when the distance $h$ between the near-field and far-field planes is relatively
small (and cannot be approximated to be infinite) while the pattern angular
dimension is small. This approach enables us to derive an analytical formula for the far-field that is expected to show qualitative agreement with the experiment.

We assume a narrow aperture width (defined as the difference between its outer and inner radii), so that the modes consist of a single lobe along the radial direction. This assumption is well justified, as will be shown in Section \ref{sec_experiment}, where the aperture in ring-VCSEL is demonstrated to possess a sufficiently small width to support the formation of single-lobed radial modes. Furthermore, we neglected the ring's width compared to
the radius $r_0$ of the ring. In the fabricated devices analyzed in this paper
their radii are $0.5\,$mm. Using the above approximations, we can describe
a single mode near-field distribution in polar coordinates as follows:
\begin{equation}
  \label{eq:mode_nf}
  \mathrm{NF}_n(r, \varphi) = A_n\delta(r-r_0)\cos(n\varphi + \alpha_n)
\end{equation}
where $n=0,1,2,\dots$ enumerates the modes, $A_n$ is the mode's amplitude,
$\delta$ is the Dirac delta, and $\alpha_n$ is a parameter
which describes the alignment of the azimuthal lobes with respect to the
abscissa axis.
In the experiment described in Section \ref{sec_experiment} we measure the far field intensity (without the information on the phase of the field). Assuming the near-field pattern given by
Eq.~\eqref{eq:mode_nf}, after calculations described in Supplementary Note 1,
we obtain the following formula for the intensity of the far field of a~single mode:
  \begin{multline}
  \label{eq:mode_ff}
  |\mathrm{FF}_n|^2(r,\varphi) = \frac{r_0^2k_n^2|A_n|^2}{h^2}J_n^2\left(k_nr_0\frac{r}{h}\right)\cos^2(n\varphi + \alpha_n) \\
    = \frac{r_0^2k_n^2|A_n|^2}{h^2}J_n^2\big(k_nr_0\tan\vartheta\big)\cos^2(n\varphi + \alpha_n)
\end{multline}
where $k_n$ is the wavenumber of the $n$-th mode, $J_n$ are the Bessel
functions of the first kind, and $\vartheta$ is the angular distance
to the vertical axis.

\begin{figure*}[bth]
  \parbox{\kol}{{\small$n=10$}\\
  \includegraphics[width=\kol]{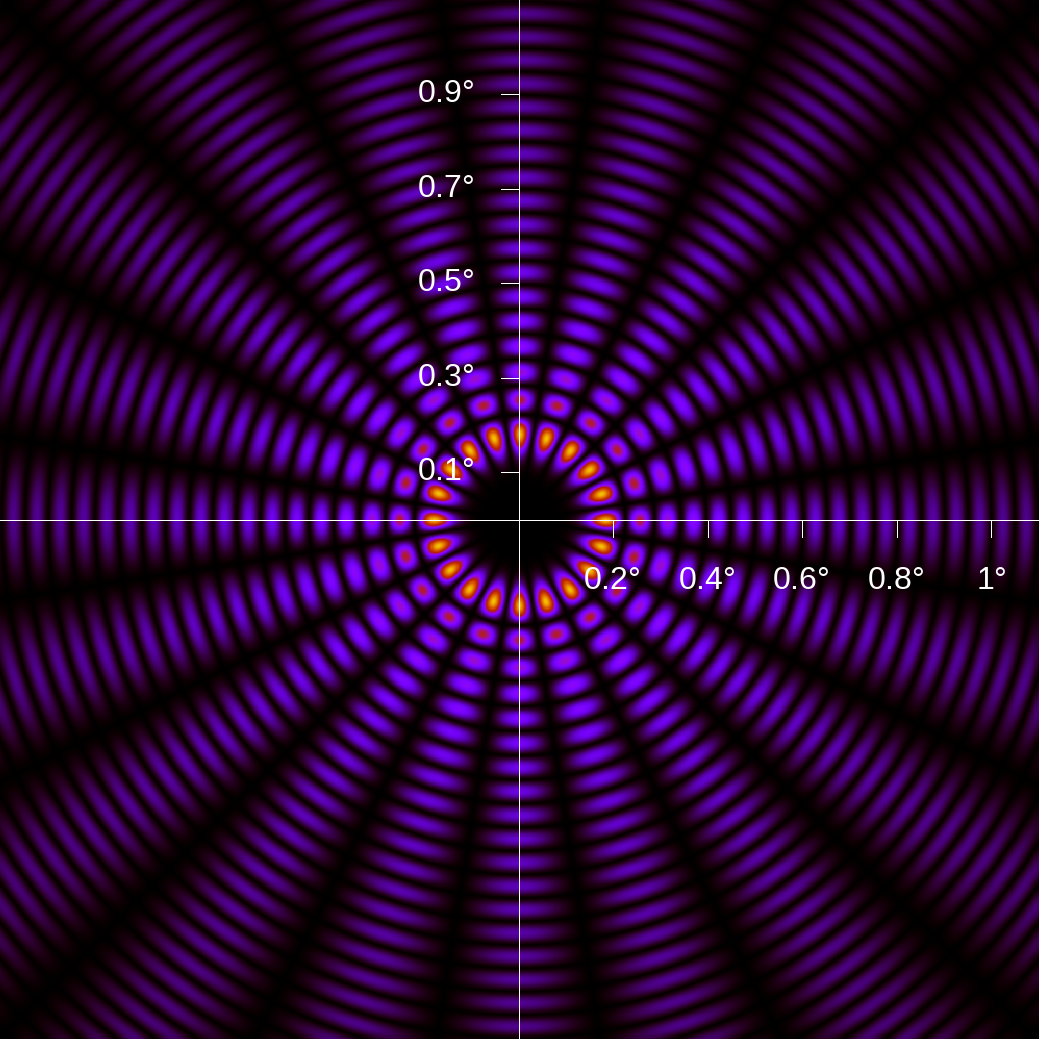}}\hfill\parbox{\kol}{{\small$n=25$}\\
    \includegraphics[width=\kol]{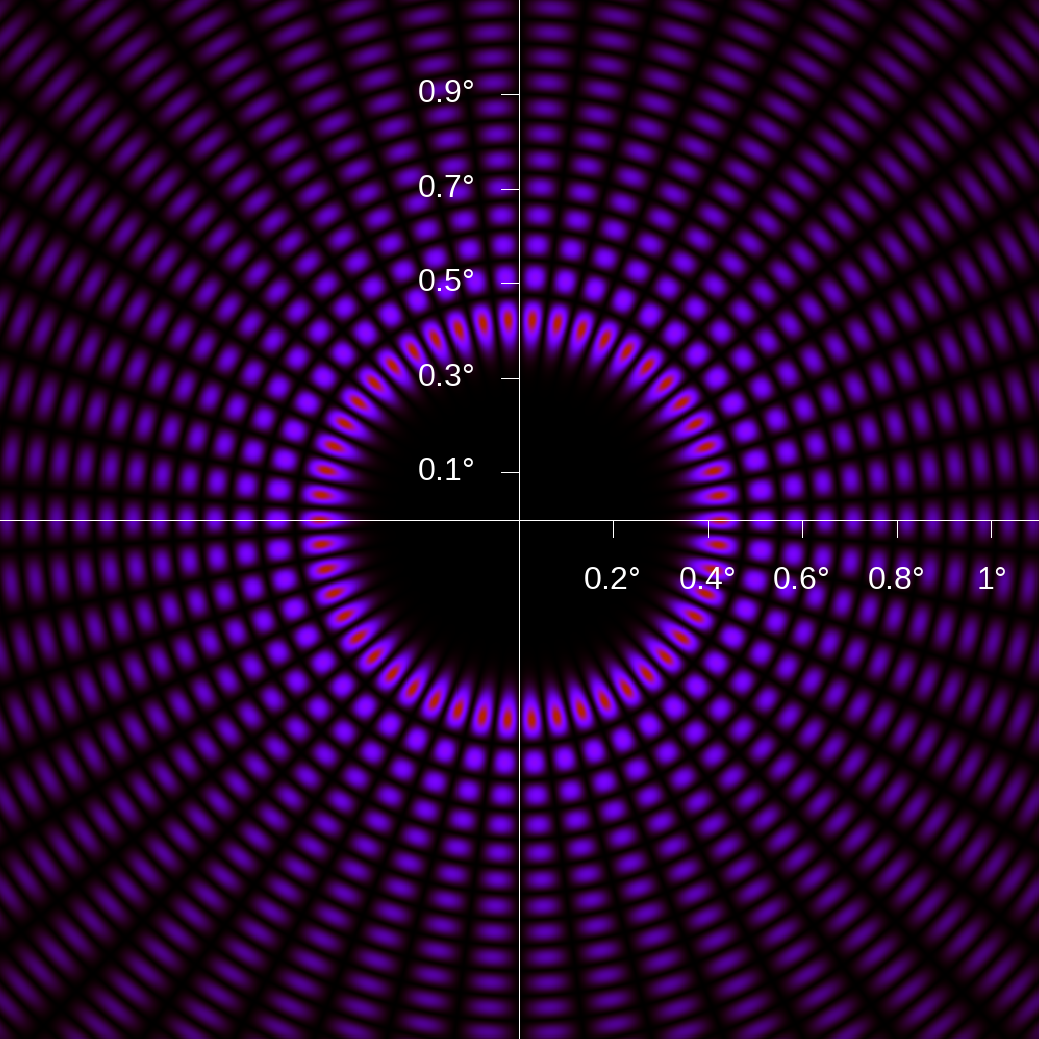}}
  \caption{Far-field patters for a mode of $n=10$ (left) and $n=25$ (right).
  The amplitudes for both modes are the same and $\alpha_{10}=\alpha_{25}=0$.
  The tics on the axes show values of angle~$\vartheta$. The assumed wavelength
  for both modes is $850\,$nm.}\label{fig:mode}
\end{figure*}

In figure~\ref{fig:mode} the far-field patterns for a mode with $n=10$ and
$n=25$ are presented.
The single-mode far-fields preserve the number ($2n$) and angular
positions of the azimuthal lobes. In the radial distribution,
there is a very characteristic dark disk in the center. The brightest
part of this pattern
is the ring which encircles the dark disk. Further from the center,
the intensity of the field slowly decreases, however the decrease rate is
underestimated by the approximation we use as will be discussed further.

The radius of the brightest ring (which determines also the radius of the central dark disk) is given by the following formula:
\begin{equation}
  \label{eq:rdd}
  R_n = \frac{\zeta_nh}{k_nr_0}
\end{equation}
where $\zeta_n$ is the position of the first (and global) maximum
of $J_n$. The values of $\zeta_n$ can be found numerically~\cite{maxbessel}.
The sequence $(\zeta_n)$ is increasing approximately linearly with increasing
$n$.
For a fixed wavelength, $R_n$ is inversely proportional
to the aperture radius $r_0$, which explains why it is possible
to obtain a very low beam divergence of a beam generated by a
very large ring VCSEL. The above formula also shows how the beam
divergence can be adjusted.

\begin{figure}[h]
  \centering
  \includegraphics[width=\kol]{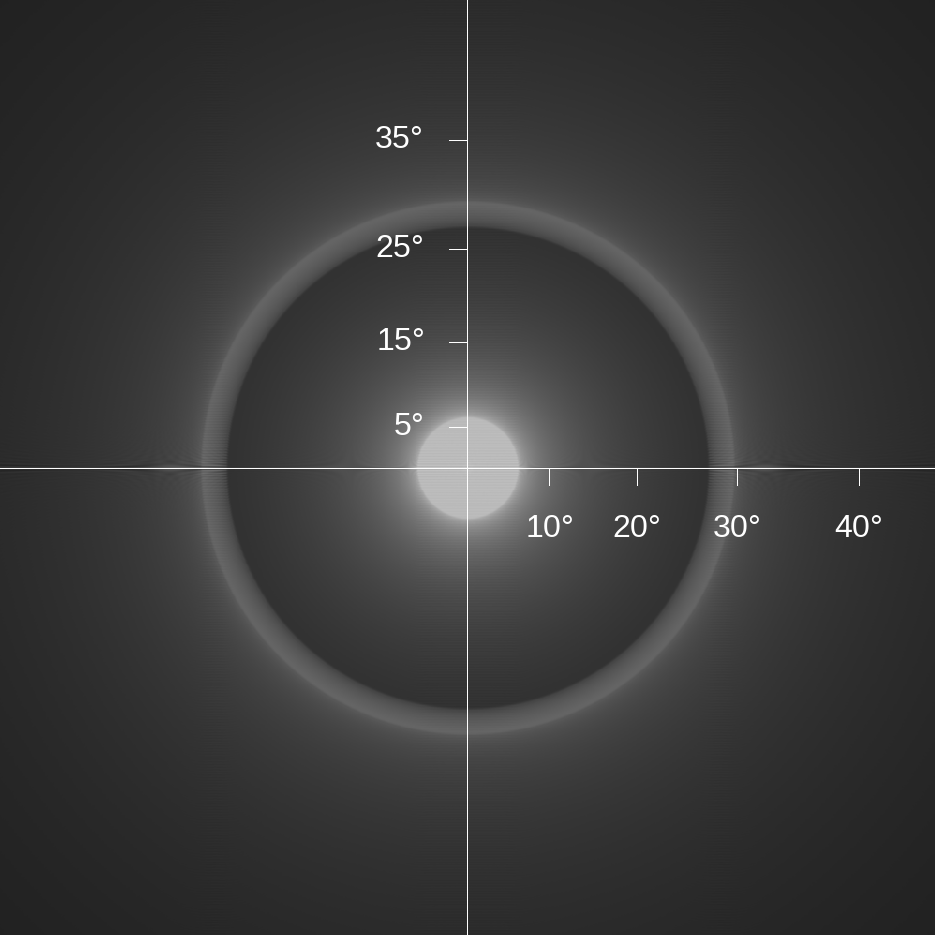}
  \caption{A far-field pattern for emission consisting of two groups
  of modes: the first containing modes with $n=0,1,\dots,400$, the
  second with $n=1900,1901,\dots,2100$. The central disk is
  generated by the lower-order modes, the outer ring by the
  higher-order group.}\label{fig:modes}
\end{figure}

One can expect that large diameter ring-VCSEL operate on large number of angular modes simultaneously. Each mode has a slightly different
wavelength, so the modes are not mutually coherent. As a result,
in the far field intensities (i.e.~squared absolute values) should
be summed to represent the overall light intensity.
Figure~\ref{fig:modes} shows a far-field pattern generated by
the modes of indices $n$ from $0$ to $400$ and from $1900$ to $2100$.
The angles $\alpha_n$ had random values for each $n$. The wavelength
of all the modes were assumed to be equal to $850\,$nm. Naturally, we expect the wavelength to decrease slightly with $n$, and we presume the associated variation to be insufficient to affect the generated far-field patterns, so we omit it in our calculations. Also all the amplitudes were assumed to be equal,
which also is an arbitrary choice.
The pattern in Figure~\ref{fig:modes} shows that the complicated structures
of the far fields of individual modes presented in Figure~\ref{fig:mode}
disappear where multiple modes of similar orders lase simultaneously.
As a result, we get patterns which are homogeneous in the angular
direction and resembles the patterns measured experimentally.

It is worth noting that the Fresnel approximation and the fact
that we neglected the width of the ring make our model underestimate
the rate of decay of the intensity when the distance from the center
increases. Let us estimate the range in which this approximation is valid.
Let us denote the width of the aperture by $\Delta R$ and
consider a point on the far-field plane whose polar-angle co-ordinate is $\vartheta$. 
At first we restrict our
considerations to the vertical plane
containing that point and the vertical axis. 
Assuming that
$\Delta R\ll h$ (which definitely seems to be the case) and that the distance
of the far-field point to the axis is much grater than $r_0$ (and hence
than $\Delta R$) we can express the difference
between the optical lengths of the beams emitted from the inner and outer
edges of the aperture separated by a distance equal to $\Delta R$ in the
following way:
\begin{equation}
  \label{eq:dl}
  \Delta l = \Delta R\sin\vartheta
\end{equation}
If $\Delta l$ is comparable to a multiple of the half of the wavelength, we expect
the beams emitted from that part of the aperture to interfere destructively
(assuming a uniform, or symmetric, mode intensity and phase distribution
along the width of the ring---which is not necessarily exactly the case).
This situation is equivalent to the determination of the position of the
minima in the single-slit diffraction. Similarly, even if the condition of being an integer multiple of half the wavelength is not satisfied, interference can still lead to a significant reduction in the light intensity compared to the value predicted by Eq.~\eqref{eq:mode_ff}, which, in the single-slit analogy, corresponds to the central peak intensity. 

If we consider other vertical planes which intersect the aperture and the
far field point under consideration, the resulting optical length
difference are larger than $\Delta R\sin\vartheta$, so we can
assume that formula~\eqref{eq:mode_nf} should be valid if
\begin{equation}
  \sin\vartheta < \frac{\lambda}{2\Delta R}
\end{equation}
Assuming $\lambda = 850\,$nm and
$3\,\text{\textmu m}\leq\Delta R\leq5\,\text{\textmu m}$
we obtain the respective condition: $\vartheta\lesssim 5^\circ$
and $\vartheta\lesssim 8^\circ$. This condition refers
to range of angles, where we should expect quantitative validity
of the model. For larger angles our model should underestimate
the radial decay rate of the optical field and overestimate
the intensity of the field (relative to the values
predicted for low angles). However, the predicted position of the
maximum of the field even for the high-order modes should be
much more reliable.
\section{VCSEL structure and experimental set up}\label{sec_experiment}
The coherent ring VCSEL (CR-VCSEL) structure developed at Dallas Quantum Devices is a monolithic high-power device with a narrow linewidth. The epitaxial design follows a conventional approach, incorporating standard Bragg reflectors and current confinement provided by an oxide aperture of approximately $4\,$µm in width. The aperture has a ring geometry with an overall diameter of about $1\,$mm.
\begin{figure*}[tbh!]
  \centering
  \includegraphics[width=2\kol]{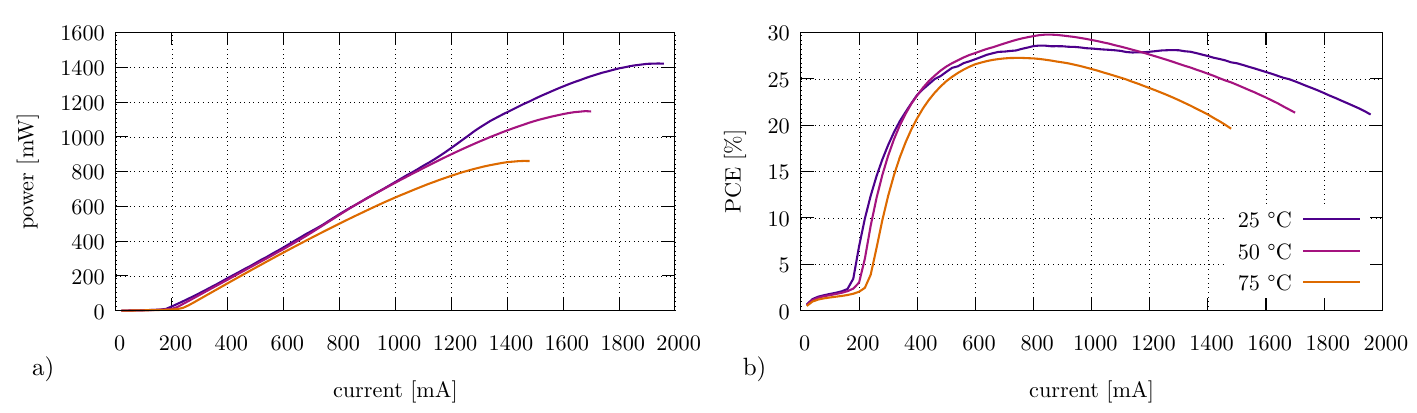}
  \caption{a) Output power measured with an integrating sphere as a function of drive current for heatsink temperatures of 25~$^\circ$C, 50~$^\circ$C, and 75~$^\circ$C. b) Corresponding power conversion efficiency (PCE) as a function of drive current for the same temperatures.}\label{fig:pce}
\end{figure*}
The output power and power conversion efficiency (PCE) characteristics measured with an integrating sphere at heatsink temperatures of 25~$^\circ$C, 50~$^\circ$C, and 75~$^\circ$C are shown in Fig.~~\ref{fig:pce}. The device exhibits output powers exceeding 1.4~W at a rollover current of 2~A, with a threshold current of 0.16~A at room temperature. Increasing the temperature leads to a gradual reduction of the emitted power at rollover to 0.86~W and an increase in the threshold current to 0.23~A at 80~$^\circ$C. The corresponding PCE reaches values close to 30\% at room temperature and decreases to 27\% at 80~$^\circ$C at moderate injection currents. These results confirm that the device operates in the watt-class regime over the investigated temperature range.
\subsection{Far-field patterns}
\begin{figure*}[t]
  \includegraphics[width=0.31\textwidth]{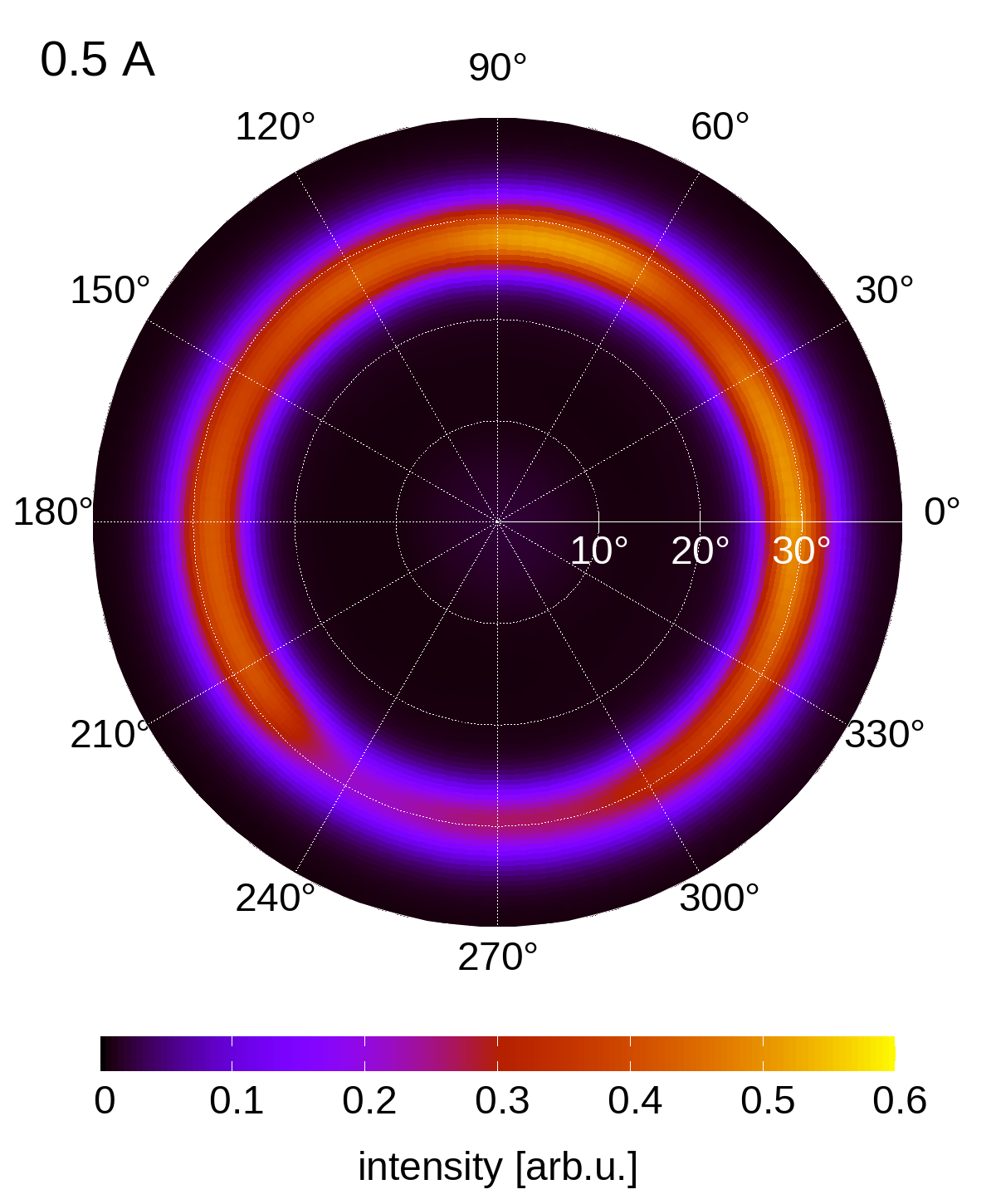}\hfill\includegraphics[width=0.31\textwidth]{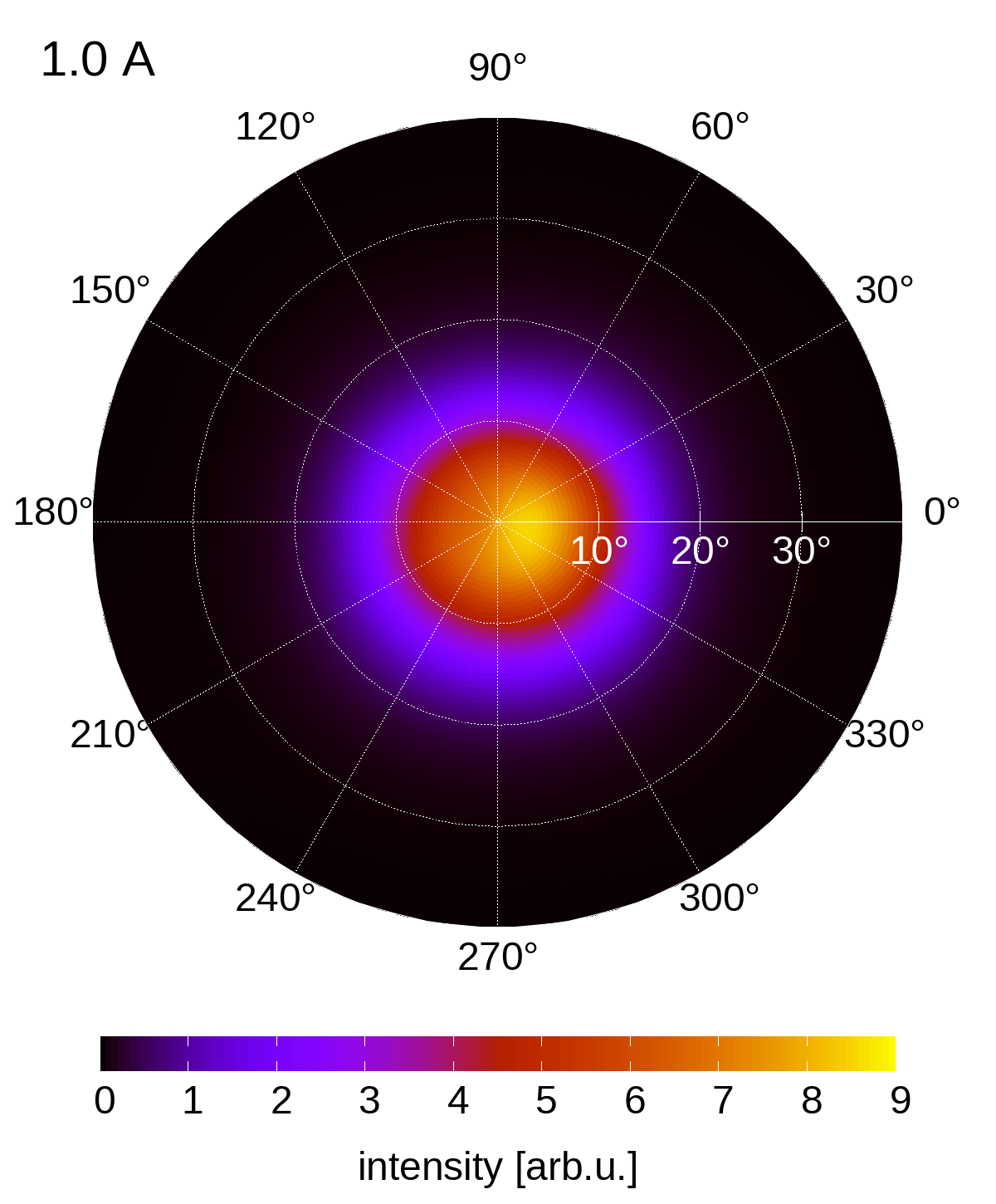}\hfill\includegraphics[width=0.31\textwidth]{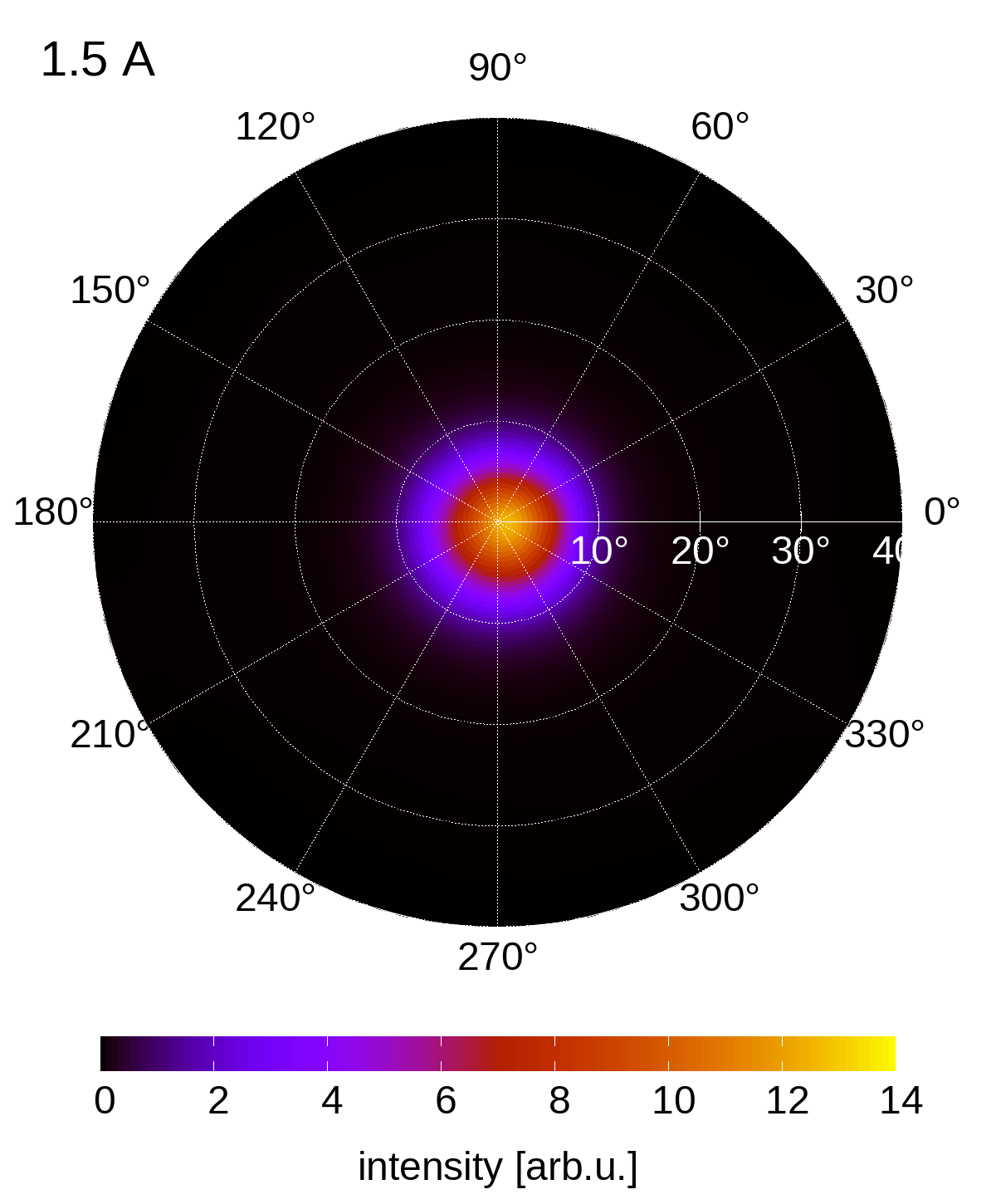}
  \caption{Far fields at different driving currents ($0.5\,$A, $1.0\,$A
    and $1.5\,$A from left to right). The unit for intensity is identical
    on all the plots, and the numbers at the color scales represent the
    actual relative intensity.}
  \label{fig:ffmap}
\end{figure*}

Using the experimental setup described in the materials provided in Supplementary Note 2, we measured far-field emission patterns of the VCSEL at several driving currents. In Figure~\ref{fig:ffmap},
high-resolution far-fields at currents: $0.5\,$A, $1.0\,$A and $1.5\,$A
are presented.
At $0.5\,$A the central disk is very weak and most of the power is
emitted in the ring of around $30^\circ$ divergence.
At a current of $1\,$A, the ring disappeared and almost the whole emitted
power is confined in the central disk. With a further current increase,
the beam becomes narrower and at $1.5\,$A and its full width at
half-maximum (FWHM) divergence angle is as low as around $8^\circ$.
The cross-sections of that beam fit very well Gaussian distribution
with $1/e^2$ diameters between $13^\circ$ and $14.5^\circ$ depending
on the cross section direction. Such low values of divergence are
characteristic for single-mode (and hence low-power) VCSELs~\cite{Peng2020},
while the output powers of the investigated laser, exceeding $1\,$W
are achieved in multi-pixel VCSEL arrays~\cite{Miller2001}.

\begin{figure*}[tbh]
  \includegraphics[width=0.24\textwidth]{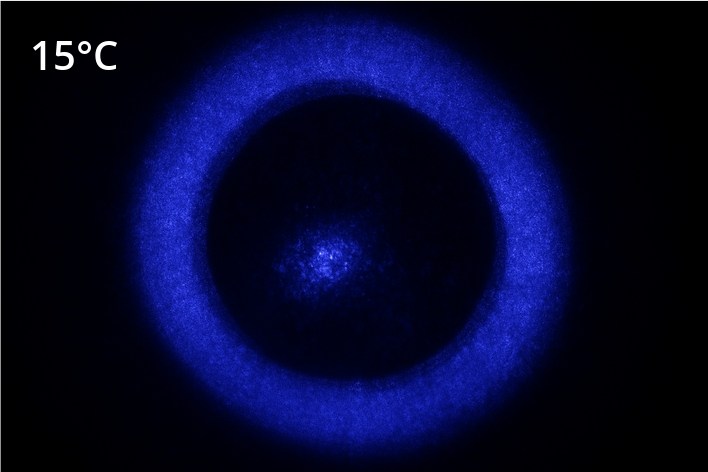}\hfill\includegraphics[width=0.24\textwidth]{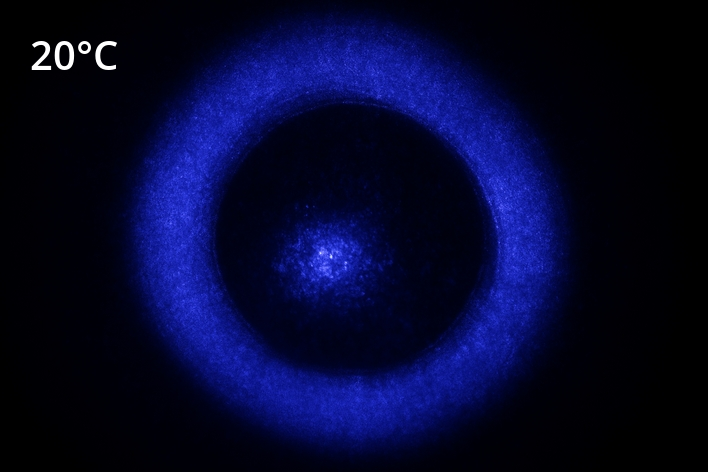}\hfill\includegraphics[width=0.24\textwidth]{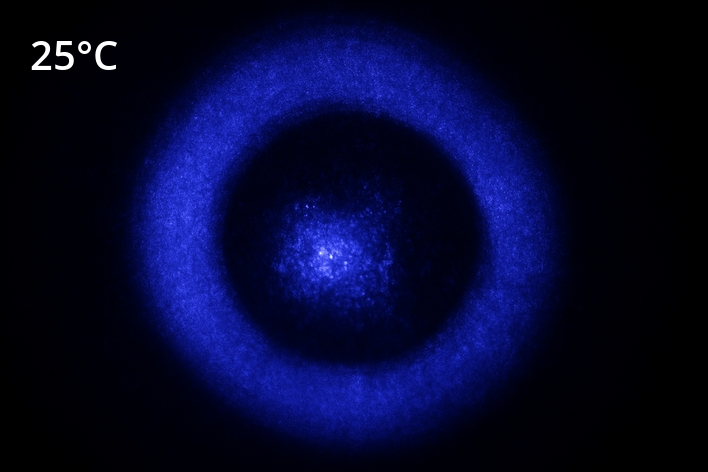}\hfill\includegraphics[width=0.24\textwidth]{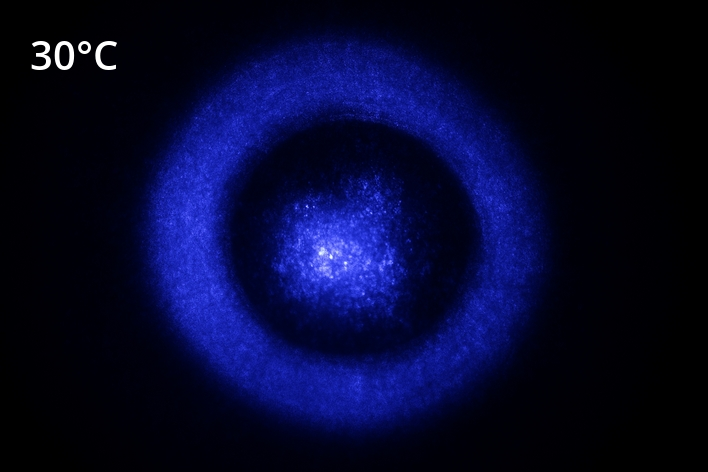}\\[0.2cm]
    \includegraphics[width=0.24\textwidth]{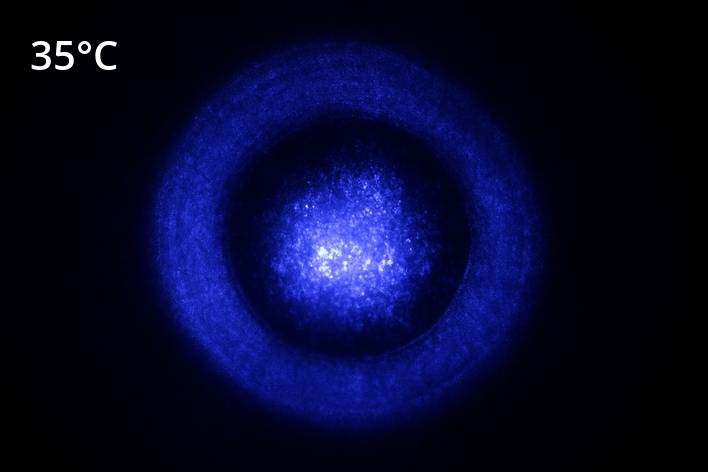}\hfill\includegraphics[width=0.24\textwidth]{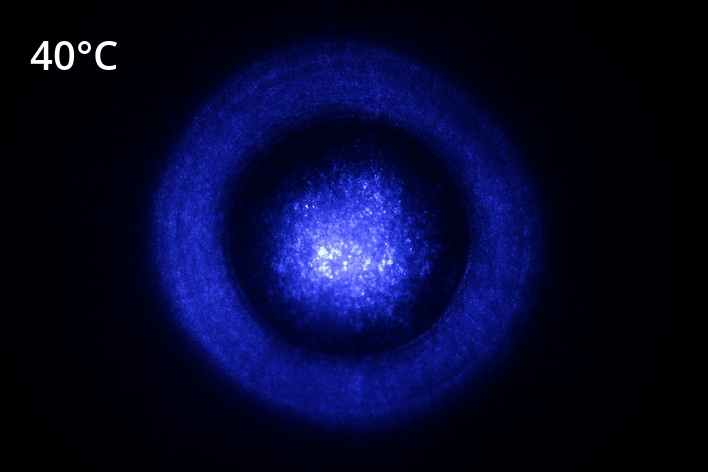}\hfill\includegraphics[width=0.24\textwidth]{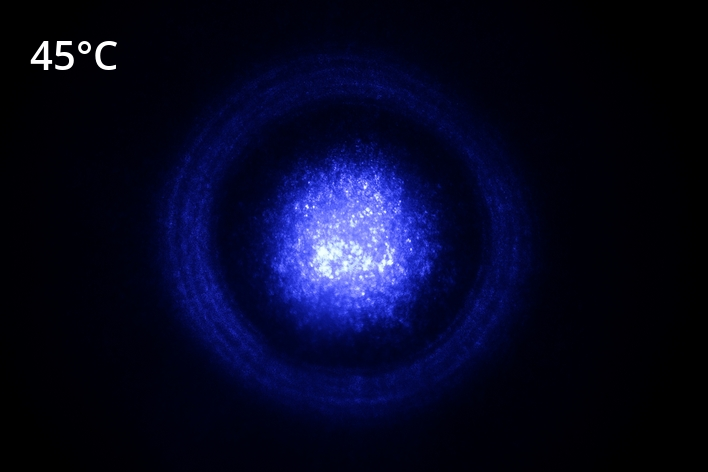}\hfill\includegraphics[width=0.24\textwidth]{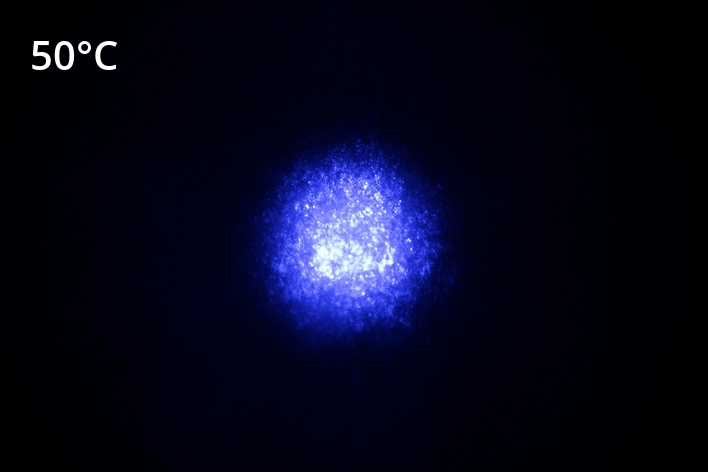}
    \caption{Far-field maps collected using the simplified method at
      a range of heat sink temperatures (indicated on each picture). The
      driving current was $0.8\,$A. The exposure parameters for the
      camera were the same for all the pictures. The brightest (white)
      pixels are overexposed.
    }
  \label{fig:fftemp}
\end{figure*}

The maps presented in Figure~\ref{fig:ffmap} show how strongly
the laser's far field depends on the operating condition. The results
presented there do not allow us to tell whether this is a thermal
effect or a result of increased carrier injection.
It is important to note that during the measurements the VCSEL operated without active heat-sink temperature stabilization. Using the setup employed to obtain the far-field maps shown in Figure~\ref{fig:ffmap}, the acquisition of a single high-quality map required more than two hours, during which temperature and other parameters could vary. This can be a reason why the ring in Figure~\ref{fig:ffmap}
does not show a good rotational symmetry (unlike in Figure~\ref{fig:fftemp}).  However, this does not affect the main result presented in this section, namely the observation of a low-divergence beam.
In order to investigate the impact of the temperature the laser operates
at, we used a simplified method for far-field observation. We use a~translucent matt screen made out of tracing paper and placed it above
the laser. We photographed the far-field patterns on the screen using a
low-magnification microscope objective and an APS-C sensor camera with the
IR filter removed. This is a very quick method, however compared with the
previously described method it has a much lower dynamic range and spatial
resolution. Additionally, the spatial range is limited by the field of view
of the objective and by how close we can place screen to the laser.
Another difference is the fact that here the patterns are projected
onto a horizontal plane, while in the other set up we measure them
on a sphere.
In Figure~\ref{fig:fftemp} far-field maps obtained using this method
are shown. The driving current was $0.8\,$A and the heat sink temperatures
were varied from $15^\circ$C to $50^\circ$C. The camera exposure
parameters i.e.~exposure time and ISO setting were kept constant,
so the light intensity observed in the pictures represent
the actual emission intensity, within the dynamic range of the camera.
At higher temperatures, the camera matrix is clearly overexposed
in the area of the central disk.

Two important conclusions can be drawn from the patterns in
Figure~\ref{fig:fftemp}. First, the shape of the laser's emission
is determined by the temperature inside the laser. In the investigated device, the typically favorable emission in a~narrow Gaussian beam is observed at higher temperatures.
Second, the intensity of the central disk increases with
increasing temperature. We do not, however, observe a narrowing
of the central disk, which is very clear when the driving current
is increased (see Figure~\ref{fig:ffmap}). The apparent effect may instead result from the increased light intensity near the center of the image combined with camera oversaturation, which can create a misleading visual impression.
\subsection{Spectral measurements}
\begin{figure}[h]
  \includegraphics{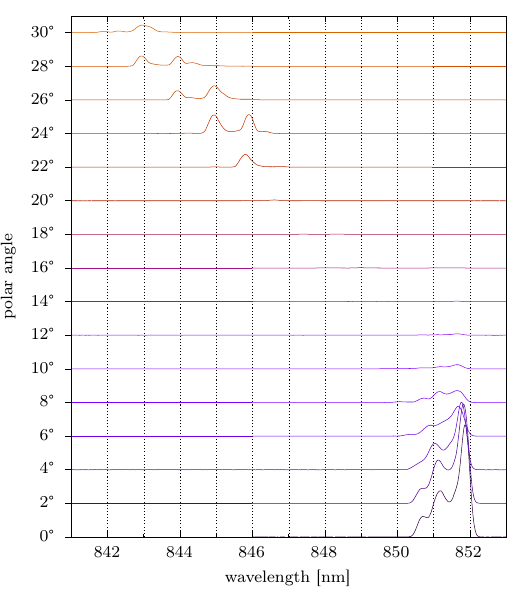}
  \caption{Polar-angle-resolved emission spectra of a laser driven by a
    current of $0.8\,$A. The spectra are in scale, i.e.~the heights of the
    peaks in all the spectra reflect their actual intensity in the same
    unit in every case. The tics at the vertical axis denote the 0 intensity
    for the corresponding spectrum.}
  \label{fig:spodteta}
\end{figure}

The theory presented in section~\ref{sec:theo} predicts that the central
disk observed in the far field should contain radiation of the low-order
(long-wavelength),
while the outer ring the higher order modes (short-wavelengths).
Using our experimental set-up
we are able to measure spectra at arbitrary points of a far-field sphere.
Since the far fields of the investigated laser are approximately
symmetric with respect to azimuthal rotations, it is enough to
investigate how the spectra vary as functions of the polar angle.
In Figure~\ref{fig:spodteta} a set of spectra indexed by the polar angle
at which they were measured is presented.
For the purpose of this measurement we set the driving to $0.8\,$A, in order
to obtain a far field containing a central disk and a ring. The spectra
clearly show that the central disk consists of the modes of the lowest orders
(the longest wavelengths), while the ring contain shorter-wavelength
radiation (higher-order modes). The dark area between the ring and the
disk corresponds to a gap in the emission spectra between roughly
$846.5\,$nm and $850\,$nm. These observations confirm at least qualitatively
the prediction of the analytical model presented in Section~\ref{sec:theo}.
It is despite the fact this model is based on certain approximation which
weaken its reliability especially for larger angles.
\subsection{Emission spectrum shaping}
In this section, we will try to present a possible reason for the gap in the
spectra presented in Figure~\ref{fig:spodteta}.
In particular, the measured far field comprises modes
spanning nearly \SI{9}{\nano\meter} with a~\SI{4}{\nano\meter} gap, and the short-wavelength contribution progressively vanishes with 
increasing temperature. Two factors are expected to play a dominant role in shaping the laser emission spectrum: (i) the emission 
spectrum of the active region and its temperature dependence, and (ii) the wavelength-dependent photon lifetime of the cavity modes. In 
the following, we determine the former experimentally and estimate the latter numerically.

\begin{figure}[tbh]
    \centering
    \includegraphics{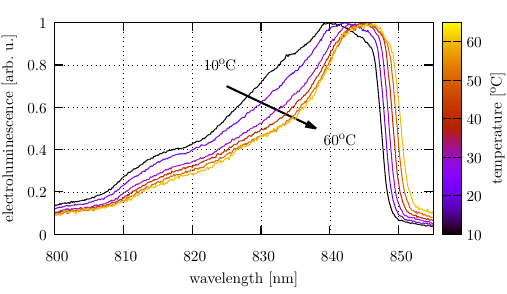} 
    \caption{Edge-emitted electroluminescence spectra of the active region measured at heatsink temperatures ranging from $10^\circ$C to $60^\circ$C in $10^\circ$C steps.}
    \label{fig:EL_a}
\end{figure}
To probe the electroluminescence of the active region with minimal influence from the vertical cavity resonance, we prepared a cleaved 
access to the epitaxial structure using a focused ion beam (FIB). A nearby electrical top contact was used to inject carriers into the 
active region, enabling direct emission in a direction parallel to the epitaxial layers and thus minimally affected by the cavity. 
Figure~\ref{fig:EL_a} shows electroluminescence spectra collected at a drive current of \SI{0.1}{\ampere} for six heatsink temperatures 
between \SI{10}{\celsius} and \SI{60}{\celsius}. The low injection current is essential to suppress the contribution of stimulated 
emission in the vertical cavity, which would otherwise distort the spontaneous emission spectrum. Owing to the reduced current density, 
the spectra in Figure~\ref{fig:EL_a} are blue-shifted relative to the lasing emission reported in Figure~\ref{fig:spodteta}. With 
increasing temperature, the electroluminescence peak systematically red-shifts and the spectrum narrows: the linewidth decreases from 
\SI{8.5}{\nano\meter} to \SI{6.5}{\nano\meter} when evaluated at \SI{80}{\percent} of the peak intensity.

\begin{figure*}[tbh!]
  \includegraphics[width=\kol]{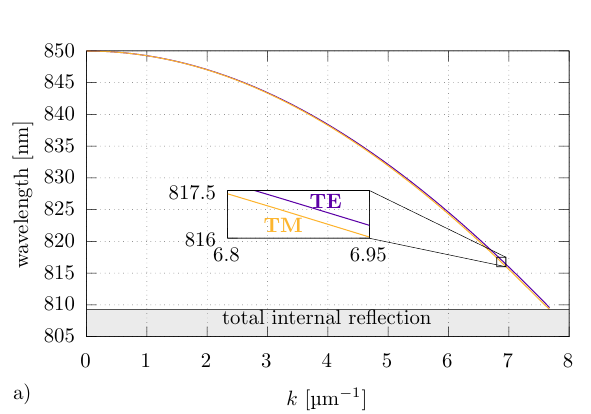}\hfill\includegraphics[width=\kol]{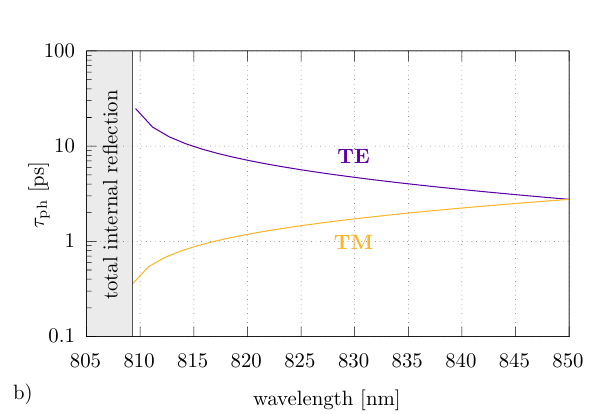}
  \caption{a) Calculated real part of the resonant wavelength as a function of the in-plane wave vector $k$ for TE and TM polarizations. The inset highlights the splitting between TE and TM polarisations. The shaded region indicates the total internal reflection regime. b) Photon lifetime $\tau_{\mathrm{ph}}$ plotted as a function of wavelength for TE and TM polarizations. The shaded region marks the total internal reflection regime.}\label{fig:EL_Q}
\end{figure*}
To calculate the photon lifetime of the cavity modes, we employ the plane-wave admittance method described in Ref.~\cite{Dems2005}, 
which provides an efficient numerical solution of Maxwell's equations for layered resonant structures. A full three-dimensional 
electromagnetic analysis of the complete device is not feasible due to its exceptionally large lateral dimensions, which would require a 
number of basis functions and computational resources beyond practical limits. We therefore adopt a~one-dimensional approach, treating 
the structure as laterally infinite, which yields the complex resonant wavelength $\tilde{\lambda}=\lambda+i\lambda_i$ as a function of 
the in-plane wave-vector component $k$. In these calculations, the refractive indices are assumed to be real; therefore, internal 
absorption and optical gain are neglected.
The resonant wavenumber is given by
\begin{equation}
k_0 = \sqrt{k^2 + k_z^2},
\label{eq:k0_relation}
\end{equation}
where $k_0 = 2\pi/\lambda$ and $k_z$ is its out-of-plane (vertical) component. The value of $k_z$ is set by the layered VCSEL structure and is assumed to be constant.

Figure~\ref{fig:EL_Q}a shows the resulting real part of the resonant wavelength as a function of $k$. Both TE and TM polarizations exhibit similar trends, with a small splitting that increases with $k$. Here, TE refers to the polarization with the electric field perpendicular to $k$, whereas TM corresponds to the polarization with the magnetic field perpendicular to $k$.

The photon lifetime $\tau_{\mathrm{ph}}$ is derived from the imaginary part of the complex resonant wavelength $\lambda_i$ according to
\begin{equation}
\tau_{\mathrm{ph}} = -\frac{\lambda^{2}}{4\pi c\,\lambda_i}.
\label{eq:tau_ph}
\end{equation}
The resulting dependence is shown in Figure~\ref{fig:EL_Q}b. Notably, decreasing the resonant wavelength (corresponding to increasing~$k$) increases the photon lifetime for TE polarization, while it decreases for TM, which can be attributed to the polarization-dependent angular reflectivity governed by the Fresnel equations. This implies that for low-order modes both polarizations may contribute comparably to stimulated emission, unless additional discrimination is imposed by the quantum-well gain anisotropy or by the cavity geometry. In contrast, at shorter wavelengths the TE polarization is favored due to its substantially higher DBR reflectivity at larger incidence angles (and thus longer photon lifetimes), which effectively lowers the material gain required at threshold.

The magnitude of $k$ corresponds to the mode number $n$, as defined in Eq.~\eqref{eq:mode_nf}, through the relation $k p = 2 n \pi$, where $p$ denotes the perimeter of the aperture. The maximum attainable value of $k$ prior to reaching the total internal reflection limit is approximately $k_{\text{max}} \approx 7.8\,\text{\mi m}^{-1}$. This imposes an upper bound on the mode order that can be emitted through the surface, yielding $n_{\text{max}} = 3900$. The mode numbers $n$ employed in the simulations shown in Figure~\ref{fig:modes} remain below this threshold.

The analysis shows that the increase in $\tau_{\mathrm{ph}}$ at shorter wavelengths (Figure~\ref{fig:EL_Q}b) may partially compensate for the reduced electroluminescence in this spectral range (Figure~\ref{fig:EL_a}), thereby enabling stimulated emission of higher-order modes that likely give rise to the outer ring observed in the far field. The temperature-induced red-shift of the gain spectrum, together with the spectral narrowing observed in Figure~\ref{fig:EL_a}, can explain the suppression of these higher-order modes at elevated temperatures, whether caused by an increased heatsink temperature or by Joule heating at higher injection currents. This observation further suggests that an appropriate initial detuning between the luminescence spectrum and the cavity resonance could be exploited to design large-aperture ring VCSELs that deliver a Gaussian-like far-field beam over a broad current range.
\section{Conclusions}
In this work, we investigated the far-field emission of a large (\SI{1}{\milli\meter}-diameter) ring-aperture VCSEL and developed a theoretical framework that captures the key features of its angular radiation patterns. Assuming an azimuthally modulated ring distribution for the near field and evaluating the far field within the Fresnel approximation, we derived analytical expressions showing how a~Gaussian-like far-field intensity profile can emerge from a large-area ring-aperture VCSEL through the contribution of numerous lower-order azimuthal modes.

Experimentally, we observed a pronounced evolution of the far-field distribution with operating conditions. At low currents, the emission is dominated by a high-divergence ring, whereas increasing the drive current progressively concentrates the optical power into a narrow central beam. At high currents, the far field approaches a near-Gaussian profile with a full width at half maximum of \SI{8}{\degree} while maintaining watt-class output power. Notably, this behavior occurs in a large-area device supporting a large number of transverse modes, indicating that a quasi-Gaussian far field can emerge under multimode operation.

Angle-resolved spectroscopy further links the central emission to longer-wavelength, lower-order modes and the outer ring to shorter-wavelength, higher-order contributions, in agreement with the theoretical model. Combined with electroluminescence measurements and numerical estimates of the wavelength-dependent photon lifetime, these results provide a consistent picture of how gain shaping and cavity losses govern the wavelength--angle emission landscape. Beyond indicating that the gain spectrum should be appropriately shifted toward longer wavelengths relative to the cavity resonance, this framework offers practical guidance for engineering divergence and mode content in large-area VCSELs, with relevance to high-power applications requiring near-Gaussian free-space beam profiles.

\section*{Competing interests}
The authors declare that they have no competing interests.

\section*{Data availability}
Relevant data supporting the key findings of this study are available within the article and the Supplementary Information file. All raw data generated during the current study are available from the corresponding author upon reasonable request.

\bibliographystyle{unsrt}
\bibliography{dallas}
\end{document}